# Development of All-Diamond Scanning Probes Based on Faraday Cage Angled Etching Techniques

C. Giese, P. Quellmalz and P. Knittel

*Fraunhofer IAF, Fraunhofer Institute for Applied Solid State Physics, Freiburg/Germany*

*We are proposing a novel fabrication method for single crystal diamond scanning probes for atomic force microscopy (AFM), exploiting Faraday cage angled etching (FCAE). Common, oxygen-based, inductively coupled plasma (ICP) dry etching processes for diamond are limited with respect to the achievable geometries. The fabrication of freestanding micro- and nanostructures is therefore challenging. This is a major disadvantage for several application fields e.g., for realizing scanning magnetometry probes based on nitrogen vacancy (NV) centres and capable of measuring magnetic fields at the nanoscale. Combining a planar design with FCAE and state-of-the-art electron beam lithography (EBL) yields a reduction of process complexity and cost compared to the established fabrication technology of micro-opto-mechanical diamond devices. Here, we report on the direct comparison of both approaches and present first proof-of-concept planar-FCAE-prototypes for scanning probe applications.*

## INTRODUCTION:

Diamond is well known for its outstanding properties such as extraordinary mechanical hardness, high thermal conductivity, wide bandgap, extremely broadband optical transparency or chemical inertness, which renders it an ideal material for various application fields [1-3]. Amongst these are single crystal-based scanning probes exhibiting excellent mechanical hardness. In recent years, it was demonstrated that colour centres in diamond could be exploited as atomically small sensors that allow for complex, highly sensitive and combined measurements in scanning probe microscopes [4].
The NV centre in diamond is the most studied among the numerous diamond point defects for applications in the fields of quantum communication, quantum computing and quantum sensing [5-9]. Combining probe tips with single NV centres at the tip apex have paved the way for highly sensitive, quantitative sensing of magnetic fields on the nanometre scale [10-12].
Over the last decade, a common fabrication method for opto-mechanical diamond devices for such NV scanning probes has been established, utilizing the thinning of commercially available diamond substrates, EBL and ICP-etching [13-15]. However, this membrane-based fabrication of diamond scanning probes is a lengthy and costly process. We therefore propose a new approach based on Faraday cage angled etching (FCAE), a technique which was developed in silicon semiconductor processing and first successfully applied to diamond in the Lončar group at Harvard University [16], especially focused on photonic applications [17, 18].

In the current study, we present our progress on the fabrication of membrane-based as well as planar FCAE all-diamond devices. Thus, we are able to compare the shown fabrication methods and will discuss the possible advantages of planar FCAE etching over the former approach. In addition, AFM measurements using scanning probes obtained via the FCAE approach are presented.

### All-Diamond Device Fabrication

The fabrication process of membrane-based vertical diamond devices can be briefly summarized as follows: Commercially available diamond membranes are nitrogen ion implanted and annealed for NV creation. An EBL-defined etch process creates vertical waveguide structures (Fig.1a I-III). In a second lithography and dry etching step, the diamond devices are isolated on the membrane. This second patterning step can either take place from the same side [13] or from the membrane backside [14,15], the latter being adopted in this work (Fig.1a IV-VI).

In contrast to this state-of-the-art process, we present a novel approach of fabricating opto-mechanical diamond devices with a planar design and FCAE. A schematic overview of the processing steps is shown in Fig.1b. Briefly, commercially available diamond substrates are wafer bonded. A single EBL and metallization step patterns the hard mask (Fig.1b I-II). Subsequently, etching is performed with a Faraday cage placed over the sample yielding freestanding nanostructures (Fig.1b III).

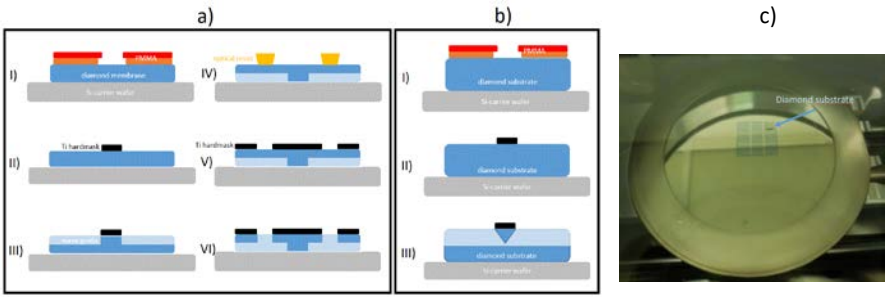

Fig.1: **a**: Flow chart of process steps for two-sided membrane-based diamond device fabrication  **b**: Process steps for FCAE-based fabrication. **c:** 3x3 mm$^2$ single crystalline diamond substrate on a silicon carrier wafer in the ICP etcher load lock chamber (Sentech SI500).

*Membrane Based Vertical Diamond Devices*

Ultrapure, single crystalline (100)-oriented diamond substrates of 3x3 mm$^2$ size and 300 µm thickness were acquired from Element Six (electronic grade, CVD). The bulk material was characterized via micro-photoluminescence (µ-PL) in order to ensure NV concentration in the low ppb range while micro Raman analysis was performed to assess the sp$^3$ phase purity and crystal quality (Renishaw Invia). Subsequently the samples were laser cut and thinned via scaife polishing to yield 30 µm thickness membranes (Delaware Diamond Knives Inc.). RMS roughness below 1 nm was verified via AFM and cleaning in a mixture of nitric acid and sulphuric acid (3:1) was performed at elevated temperatures for 120 minutes to remove graphitic and non-carbon contaminations. After nitrogen ion implantation and annealing [19], the diamond plates were temporarily bonded to 3" silicon wafers (see Fig.1c) with the implanted surface facing upwards (HT10.10 bonding agent / Brewer Science, Süss SB6). The bonding enables easier handling and use of semi-automatic tools during cleanroom processing.

The hard mask for dry-etching of vertical diamond waveguides was defined via EBL and metal evaporation followed by a metal lift-off process. For electron beam lithography a double layer of poly(methyl methacrylate) based resist (PMMA) was spin coated and covered with a 12 nm layer of evaporated aluminium to limit charging effects in the subsequent exposure step (JBX9300FS, 100 keV, JEOL). 900 µC/cm$^2$ were chosen as optimal EBL dose. After lithography, the PMMA resist was developed in a mixture of isopropanol and methyl isobutyl ketone (MIBK) (3:1) for four minutes. In a consecutive step, a 75 nm thick layer of titanium was deposited via electron beam evaporation and the underlying non-exposed resist was lifted off the diamond surface in aceton and a dimethyl sulfide jet (20 bar). Due to its low sputter yield of 0.081 (100 eV argon ions) [20] titanium is well suited as mask material for diamond structuring. Selectivity above 100 can be achieved for oxygen based dry etching processes.

Isolated, conically shaped, vertical waveguides were structured via ICP reactive ion etching (ICP-RIE / Sentech SI500). Etching parameters were as follows: 1.3 Pa chamber pressure, 700 W ICP power, 70 W RF platen power, substrate bias of -140 V and 30 sccm oxygen flow.

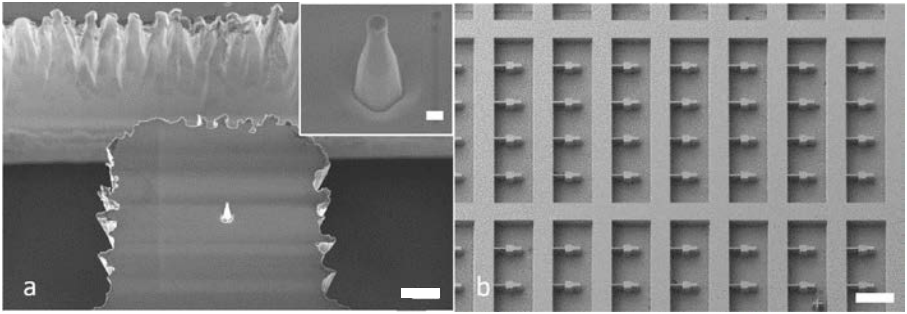

Fig.2: **a**: SEM image of vertical waveguide on isolated diamond device (scale bar 2 µm). Inset: close-up of wave guiding tip (scale bar 200 nm). **b**: SEM image of single crystalline membrane of 30 µm thickness after both etch steps: released diamond devices with single tips (scale bar 100 µm).

The achieved etch depth was roughly 3 µm at an etch rate of 200 nm/min. Fig. 2a shows a single wave-guiding tip on a finished diamond device as well as a close-up of a tip (inset). The typical geometries exhibit high aspect ratio with top diameter of 200 nm and bottom diameter of about 600 nm [18]. Using deep etching capabilities, we chose a two-sided structuring approach, similar to the approach of the Yacoby group [15]. For backside lithography, the diamonds were debonded, flipped and rebonded on the carrier wafer after removal of the metal mask in hydrofluoric acid. Image reversal i-line mask aligner lithography (Süss MA6, AZ5214 resist, 1.4 µm thickness) was performed to pattern the structures and an additional 400 µm unstructured frame around the final devices was designed to ensure mechanical stability of the membrane after deep etching. For this additional 30 µm etch step a thicker 100 nm titanium mask was evaporated to increase the possible etching time. This final etch step results in diamond opto-mechanical devices with dimensions of 20x40 µm$^2$ isolated from the membrane, remaining attached to the bulk only via a thin anchoring beam as predetermined breaking point (Fig.2b).

*FCAE Based Planar Diamond Devices*

For the proposed novel device design, the FCAE principle is shortly introduced. In low pressure reactive ion etching, a bias potential is formed between the neutral plasma and the cathode, which the sample is placed on. The latter is due to the light electrons following the capacitively coupled RF field in the MHz range applied to the electrode, charging it negatively to up to several hundred volts. The bias field accelerates the heavier ions towards the cathode where they physically etch the target material. When a metallic Faraday cage is placed over the sample, the bias field follows the geometry of the cage and the impact angle of the ions is controlled by the inclination of the cage walls (Fig.3a) [16]. Likewise, freestanding, triangular shaped diamond beams of complex geometry can be etched.

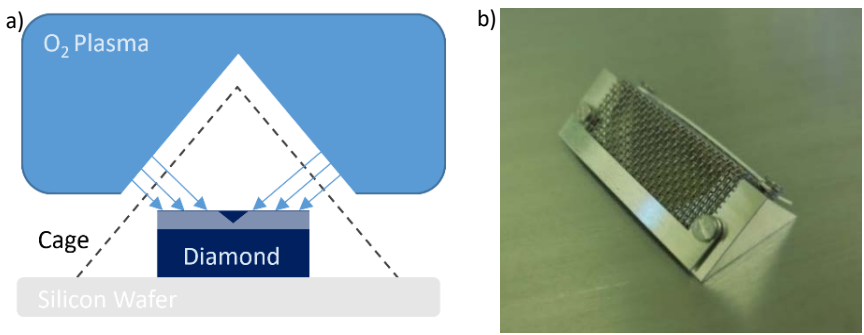

Fig.3: **a**: Schematic drawing of FCAE principle. Ions from the plasma are accelerated towards the cage surface yielding an angled incidence on the diamond substrate. **b**: Faraday cage (45° angle): aluminium frame with clamped stainless steel mesh with 240 µm wire diameter and 740 µm pitch.

For prototype fabrication 4x4 mm$^2$ high-pressure high-temperature (HPHT) IIa electronic grade crystals (New Diamond Technologies) of 500 µm thickness and (100) crystal orientation were wafer bonded as reported above (s. Fig.1,c). The nitrogen concentration is specified by the supplier to be below 10 ppb. The samples were cleaned and characterized identical to the diamond substrates used for the membrane-based scanning probes. The lithography process (Fig. 1,bI-II) was chosen identical to the membrane process (Fig. 1, a I-II). After EBL, a 150 nm nickel layer was evaporated and metal lift-off was performed. The mask material was chosen as the result of tests with a variety of materials, e.g. titanium, aluminium, nickel and SiN. Nickel exhibits the least redeposition of mask material on etch side walls, which has proven to be crucial for the FCAE process.

The Faraday cage used in the present work was a tent-like aluminium frame with clamped stainless steel meshes and sidewall angles of 45° (Fig.3b). The dimensions of the cage were 20x10x7 mm$^3$, the mesh wire diameter was 240 µm and the pitch 740 µm.

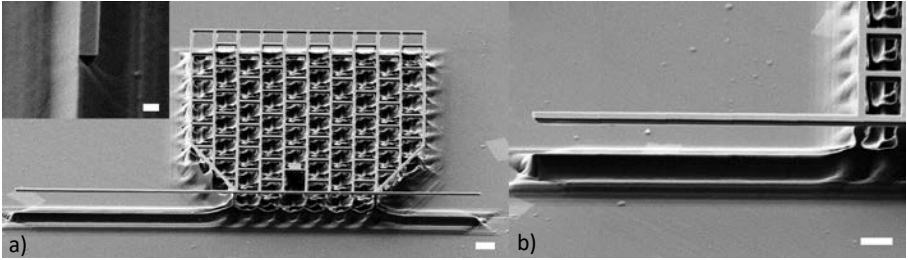

Fig.4: **a**: SEM image of scanning probe structure on the bulk diamond substrate, the grid being used for mechanical interfacing (scale bar 3 µm). Inset: Close up of triangular shaped front facet of freestanding 50 µm length beam (scale bar 400 nm). **b**: SEM close-up side view of a lateral wave guide with 20µm length (scale bar 2 µm)

FCAE was performed placing the cage over the diamond sample on the carrier wafer and ICP parameters were slightly modified as compared to the vertical waveguide etch. For further enhancement of mask selectivity, RF platen power had to be reduced to 45 W, yielding a bias voltage of -100 V. As the cage design only allows for simultaneous etching from two directions, iterative 2 minute etch steps were performed, switching between two orthogonal orientations of the cage. SEM measurements confirmed complete underetching of the target structures after 22 min for each cage orientation. Lateral etch rates of 7-10 nm/min were determined, while vertical etch rates were 110 nm/min. The total etch depth of the finished devices after 44 minutes was 4800 nm (Fig. 4a and b).

In order to enable the transfer of the devices, we defined a fine grid as mechanical interface (Fig.4a). A single beam attached to the grid acts as both, an AFM tip and a planar wave guiding structure allowing for high photon collection efficiency of NVs if located at the extremity of the triangular beam (Fig.4b). The width of the freestanding beam was 450 nm and the acute bottom angle of the triangle was approximatively 72°.

## Mounting of AFM Scanning Probes

Employing the fabricated micro-opto-mechanical diamond devices in AFM probes necessitates mechanical interfacing with quartz tuning forks (Fig.5 and Fig. 6). In order to allow for simultaneous AFM scanning and optical read-out with high numerical aperture (NA) objectives, the diamond device is attached to the tuning fork (3x1 mm$^2$) via a silicon-on-insulator (SOI) spacer with millimetre length. Fig.5a shows a SEM side view of the SOI-mounted membrane-based diamond device with a thickness of roughly 30 µm and the wave guiding tip with 3000 nm length facing down (close-up view in inset). The fork is fixed to a custom PCB with contact pads for electrical drive and read-out control, which can be mounted in the AFM head. The schematic in Fig.5b visualizes the membrane based scanning probe design.

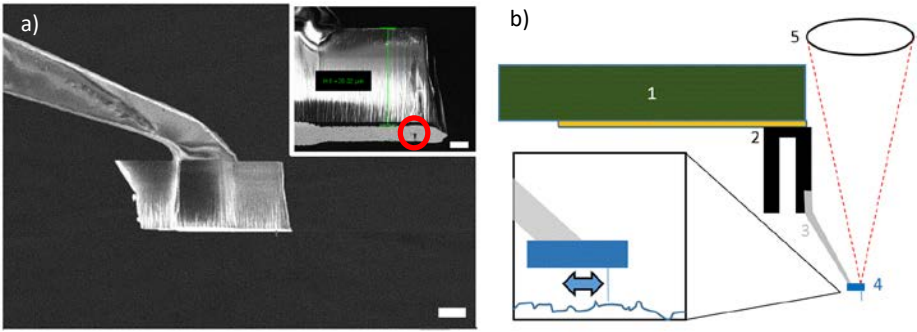

Fig.5: **a**: SEM image of mounted SOI handle and diamond micro-optic device (scale bar 10 µm). Inset: Close-up of diamond slab with highlighted 200 nm diameter vertical, conically shaped waveguide (scale bar 5 µm). **b**: Sketch of membrane-based AFM scanning probe: 1: PCB, 2: quartz tuning fork, 3: SOI spacer, 4: diamond opto-mechanical device, 5: free-space optics for colour centre excitation and read-out.

The FCAE scanning probe assembly is identical. For comparison, Fig.6a shows the diamond device (inset) attached to the SOI spacer and the tuning fork. The scanning probe design is sketched in Fig.6b. This visualizes the large similarities of the scanning probes, differing only in the diamond device geometry. Handling of the micro-diamond devices was conducted with micromanipulators (MiBot, Imina Technologies SA) and UV curing optical adhesive (NOA81, Norland) was used for fixation.

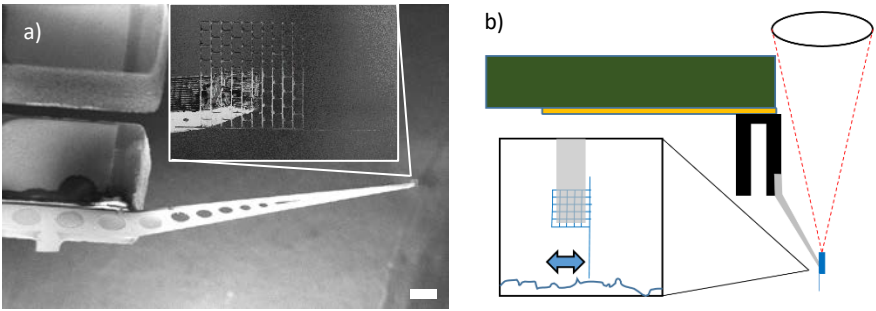

Fig.6: **a**: SEM image of quartz tuning fork with mounted SOI handle and diamond device (scalebar 100 µm). Inset: Mounted FCAE diamond device consisting of grid and triangular shaped 300 nm width beam (scalebar 2 µm). **b**: Sketch of the FCAE-based scanning probe. Components are identical to Fig. 5b. The probe is rotated 90° in the SEM image.

*Scanning Probe Characterization*

The performance of the FCAE scanning probes shown in Fig.6 were investigated in a JPK Nanowizard III AFM system equipped with a tuning fork module. A resonance frequency of approx. 32.75 kHz was chosen for the actuation of the diamond tip. Due to the attachment of the tip including the silicon handle, a decrease in resonance frequency of 0.5 kHz was measured.

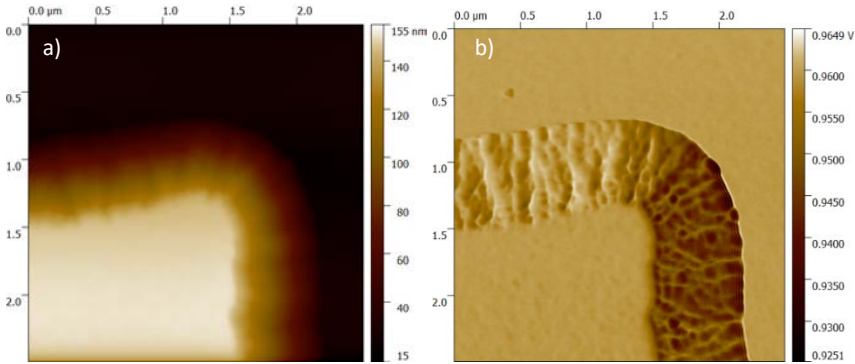

Fig.7: 2,5 µm$^2$ AFM scan of silicon oxide calibration sample with FCAE scanning probe prototype (5 nm pixel size). **a**: height image of mesa sidewall. **b**: Amplitude signal showing high resolution features of silicon oxide structure.

The ultra-high aspect ratio of the produced tip (>60) is suited for scans of target structures with steep sidewalls which are inaccessible for standard pyramidal silicon AFM tips (54° sidewall angle). Height (Fig.7a) and amplitude signals (Fig.7b) of a SiO$_x$ calibration standard mesa structure were recorded. Apart from the overall topography of the mesa, which is precisely imaged, the geometry of the sidewalls is obtained with high resolution. Especially the amplitude signal clearly reveals the sidewall morphology resulting from the fabrication process of the calibration standard.

## Discussion

FCAE-based scanning probe fabrication yields key advantages over the membrane-based technology. The scanning probe design is easily adopted in existing setups because only the diamond device is modified (Fig.5b vs. Fig.6b). Using bulk diamond substrates surmounts the challenging and costly production of fragile electronic grade diamond membranes [21,22]. Additionally, reuse of the ultra-pure substrates is possible making maximum use of the bulk material, largely lost in the case of membrane polishing. The method overcomes wedge issues encountered during membrane fabrication, readily enabling the processing of larger substrates. Using the presented approach, more than 2000 devices could be placed on a single 4x4 mm$^2$ diamond substrate, further optimization being possible. The advantages of FCAE-based diamond structuring offer new possibilities for multiple application fields. The facile fabrication of high aspect ratio (HAR) AFM tips [23,24,25] using the robust material diamond is one of them, addressing issues in AFM studies of high-aspect samples. Furthermore, the scalable technique for designing larger mechanical interface structures paves the way for automated assembly of all-diamond AFM tips employing state-of-the-art pick-and-place bonder tools. Single crystalline diamond MEMS [26,27] and photonic devices e.g. optical waveguides or resonators [28] are also promising areas of application, especially when combined with NV devices.

## Conclusion and Outlook

The presented novel FCAE-based method for producing all-diamond scanning probes opens up new design and fabrication possibilities for micron scale opto-mechanical devices. Especially the application to the scalable production of scanning probes for NV-based sensing is of interest to the growing community involved in the field of quantitative nanometre scale magnetometry.

We thank Jan Rhensius and Christian Degen from ETH Zurich for supplying SOI structures for scanning probe mounting and fruitful discussions. This work was funded within the European H2020 framework "AsteriQs" grant no. 820394 and the Fraunhofer-Max-Planck Cooperation program "DIANMR" grant no. 600717.